\documentclass[aps,pra,superscriptaddress,reprint]{revtex4-2}

\usepackage{amsmath}
\usepackage{amssymb}
\usepackage{mathrsfs}
\usepackage{graphicx}
\usepackage{booktabs}  
\usepackage{multirow}  

\usepackage[colorlinks=true, letterpaper=true, pdfstartview=FitV, linkcolor=blue, citecolor=blue, urlcolor=blue]{hyperref}

\begin{document}
	
	\title{Multiphoton blockade by multi-tone drive}

	\author{Guang-Yu Zhang}
	\affiliation{Key Laboratory of Low-Dimensional Quantum Structures and
		Quantum Control of Ministry of Education, Key Laboratory for Matter
		Microstructure and Function of Hunan Province, Department of Physics and
		Synergetic Innovation Center for Quantum Effects and Applications, Hunan
		Normal University, Changsha 410081, China}

\author{Zhi-Hao Liu}
\affiliation{Key Laboratory of Low-Dimensional Quantum Structures and
Quantum Control of Ministry of Education, Key Laboratory for Matter
Microstructure and Function of Hunan Province, Department of Physics and
Synergetic Innovation Center for Quantum Effects and Applications, Hunan
Normal University, Changsha 410081, China}
 
\author{Jie-Qiao Liao}

\affiliation{Key Laboratory of Low-Dimensional Quantum Structures and
	Quantum Control of Ministry of Education, Key Laboratory for Matter
	Microstructure and Function of Hunan Province, Department of Physics and
	Synergetic Innovation Center for Quantum Effects and Applications, Hunan
	Normal University, Changsha 410081, China}
\affiliation{Hunan Research Center of the Basic Discipline for Quantum Effects and Quantum Technologies, Hunan Normal University, Changsha 410081, China}
\affiliation{Institute of Interdisciplinary Studies, Hunan Normal University, Changsha, 410081, China}

\author{Xun-Wei Xu}
\email{xwxu@hunnu.edu.cn}
\affiliation{Key Laboratory of Low-Dimensional Quantum Structures and
Quantum Control of Ministry of Education, Key Laboratory for Matter
Microstructure and Function of Hunan Province, Department of Physics and
Synergetic Innovation Center for Quantum Effects and Applications, Hunan
Normal University, Changsha 410081, China}
\affiliation{Hunan Research Center of the Basic Discipline for Quantum Effects and Quantum Technologies, Hunan Normal University, Changsha 410081, China}
\affiliation{Institute of Interdisciplinary Studies, Hunan Normal University, Changsha, 410081, China}
	
	\date{\today}
	
	\begin{abstract}
Multiphoton blockade provides an efficient way to achieve entangled photon sources and leads to wide applications in modern quantum technologies. 
Here, we propose a scheme to realize multiphoton blockade by a multi-tone drive. 
Specifically, we demonstrate two-photon and three-photon blockades in a single-mode optical Kerr resonator using a two-tone and a three-tone drive, respectively. 
In comparison with the single-tone drive, except for the blockade of the $(n+1)$th photon excitation due to large detuning, the key mechanism in this scheme is the sequently resonant excitations of all the $m$-photon states ($m\leq n$) by the $n$-tone drive, which lead to the enhancement of photon generation and the demonstration of multiphoton blockade in the weak driving regime. 
Moreover, the photon distribution within the system can be adjusted on demand by tuning the relative amplitudes of the driving fields for different frequencies.
The scheme can be extended to other bosonic systems and be applied to demonstrate other multiphoton physical effects.

	\end{abstract}
	
\maketitle

\section{Introduction}
Photon blockade (PB)~\cite{PhysRevLett.79.1467}, a typical quantum optical effect, describes the phenomenon that the excitation of a single photon blocks the excitation of subsequent photons in coherently driven nonlinear optical systems.
The PB is considered as one of the physical mechanisms for generating single photons via coherent optical drive~\cite{Buckley_2012,RevModPhys.87.347} and single photons play a pivotal role in advancing quantum computing technologies~\cite{RN565,RevModPhys.79.135}, constructing quantum networks~\cite{RN566,RevModPhys.87.1379}, developing quantum cryptographic protocols~\cite{RevModPhys.81.1301,RevModPhys.92.025002}, and enhancing quantum sensing capabilities~\cite{RevModPhys.89.035002,Pirandola2018NaPho}.
In general, the PB can be classified into two categories: conventional PB and unconventional PB.
Conventional PB is induced by the energy anharmonicity in the optical systems with strong nonlinear interactions~\cite{Ridolfo2012PRL,LiuYX2014PRA,PhysRevA.90.033831,PhysRevLett.107.063601,PhysRevLett.107.063602,LiaoJQ2013PRA,XieH2016PRA,Majumdar2013PRB,Huang2022LPRv,Chakram2022NatPh,Zhou2020PRA}, 
and has been observed in cavity quantum electrodynamics (QED) systems~\cite{Birnbaum2005Natur,Dayan2008Sci,Aoki2009PRL,Faraon2008NatPh,LangC2011PRL,Hoffman2011PRL}.
Unconventional PB is induced by destructive quantum interference among different pathways for two-photon excitation~\cite{PhysRevLett.104.183601,PhysRevA.83.021802,PhysRevA.83.021802,Xu_2013,savona2013UPB,ZhangWZ2015PRA,Ferretti_2013,PhysRevA.88.033836,Gerace2014PRA,PhysRevA.90.043822,PhysRevA.90.033809,Lemonde2014PRA,PhysRevA.91.063808,ZhouYH2015PRA,PhysRevA.96.053810,PhysRevA.96.053827,ShenS2019PRA,Zubizarreta2020LPR,Wang_2020NJP,PhysRevLett.108.183601,ZhangW2014PRA,TangJ2015NatSR,LiangXY2019PRA,PhysRevLett.125.197402,WangY2021PRL,li2023enhancement,lu2024chiral,Andrew2021sciadv,MaYX2023PRA,ZhouYH2022PRAPP,SuX2022PRA,Ben-Asher2023PRL,ZuoYL2022PRA}, and it has been observed in both the optical~\cite{Snijders2018PRL} and microwave~\cite{Vaneph2018PRL} regimes.

The concept of PB has been extended from single-photon blockade (1PB) to multiphoton blockade~\cite{SHAMAILOV2010766,PhysRevA.87.023809}. 
Different from the 1PB that suppress the generation of more than one photon, multiphoton blockade or $n$-photon blockade ($n$PB) means that the $n$-photon state is excited without remarkable suppression, while the generation of the $(n+1)$th photon is blockaded significantly. The $n$PB can be utilized for the generation of entangled photons~\cite{yang_sequential_2022}, and
it has been studied theoretically in various configurations, such as Kerr-type nonlinear cavities~\cite{PhysRevA.87.023809,Miranowicz1996QuantumSE,PhysRevLett.121.153601,PhysRevA.104.053718,AQT2023zhang,PhysRevA.110.053702}, cavity QED systems~\cite{PhysRevResearch.6.033247,PhysRevA.102.053710,PhysRevA.95.063842,PhysRevA.98.043858,SHAMAILOV2010766,PhysRevX.5.031028,PhysRevA.91.043831,PhysRevX.5.031028,PhysRevA.98.053859,PhysRevLett.122.123604,PhysRevA.99.053850,PhysRevA.109.043702,Lin:24,PhysRevA.107.043702,NP_Hamsen,OC_Th.K.Mavrogordatos,PhysRevA.110.043707,Tang:23}, optomechanical systems~\cite{PhysRevA.103.043509,Zhai:19}, optomagnonic microcavities~\cite{PhysRevA.100.043831}, and other systems~\cite{PhysRevA.100.053857}. In particular, the 2PB has been demonstrated experimentally in an atom-driven cavity QED system~\cite{PhysRevLett.118.133604} and in a photonic-crystal nanocavity with an embedded quantum dot~\cite{PhysRevA.90.023846}.

There are two problems that need to be considered for the $n$PB when using a continuous-wave single-tone drive. Firstly, the $n$PB is driving-strength dependent and usually a strong driving strength is required to induce $n$PB~\cite{PhysRevA.87.023809}.
Secondly, the $n$PB is achieved by the resonant transition from vacuum directly to $n$-photon states, while the transitions between other photon number states are largely detuned due to strong nonlinearity, which lead to the suppression of photon generation in $n$PB.

In this paper, we propose a scheme to achieve multiphoton blockade in an optical Kerr resonator by a multi-tone drive. 
Different from the previous schemes with a single-tone drive~\cite{SHAMAILOV2010766,PhysRevA.87.023809}, here, all the transitions from the $(m-1)$-photon to $m$-photon states ($m\leq n$) in the nonlinear resonator are driven resonantly by a $n$-tone drive with appropriate frequencies, and the achievement of $n$PB is attributed to the blockade of the excitation of the $(n+1)$th photon due to the large detuning.
Thus, the $n$PB can be realized in the weak driving regime by a multi-tone drive.

The PB has also been proposed in the optical nonlinear systems driven by coherent fields with time-varying amplitudes~\cite{PhysRevA.93.043857,PhysRevA.90.013839,PhysRevA.90.063805,PhysRevB.97.241301,PhysRevLett.123.013602,PhysRevLett.129.043601,LIU2024,Geng:24,PhysRevA.110.023718} or employing time-dependent coupling~\cite{Stefanatos2020PRA}, which are referred to as dynamical blockade. Notably, dynamical blockade can be realized in weakly nonlinear systems, such as a bosonic mode coupled to two additional modes through four-wave mixing~\cite{PhysRevA.90.063805}, coupled to a gain medium~\cite{PhysRevB.97.241301}, driven by a combination of continuous and pulsed fields~\cite{PhysRevLett.123.013602}, excited by a bi-tone coherent field~\cite{PhysRevLett.129.043601}, or driven solely by a pulsed field~\cite{PhysRevA.110.023718}.
Besides, multiphoton blockade was proposed in a Kerr-nonlinear resonator driven by a sequence of Gaussian pulses~\cite{PhysRevA.90.013839}.
One problem requiring further consideration is $n$PB based on pulse drive appears periodically in the form of impulses, which limits the broader applications.

Different from the periodic $n$PB generated by the pulse-driven field~\cite{PhysRevA.90.013839}, here, we show that even though the driving amplitude of the multi-tone drive is time-dependent, multiphoton blockade can be continuously achieved in the stationary regime. 
Moreover, we can manipulate the photon number distribution by tuning the relative amplitudes of the driving field with different frequencies, and multiphoton blockade can be achieved with a relatively modest nonlinear strength.
The scheme imposes no restrictions on the number of photons for blockade and can be easily extended to other bosonic systems. 

The paper is structured as follows:
In Sec.~\ref{BIN}, we introduce the physical model of an optical Kerr resonator driven by a multi-tone drive and present the two criteria for the $n$PB effect.
In Secs.~\ref{CIN} and \ref{DIN}, we demonstrate the 2PB and 3PB in the optical Kerr resonator under the two- and three-tone drivings, respectively.
Finally, we conclude this work in Sec.~\ref{Con}.

	\begin{figure*}[t]
		\centering
		\includegraphics[bb=0 13 1879 775, width=14cm, clip]{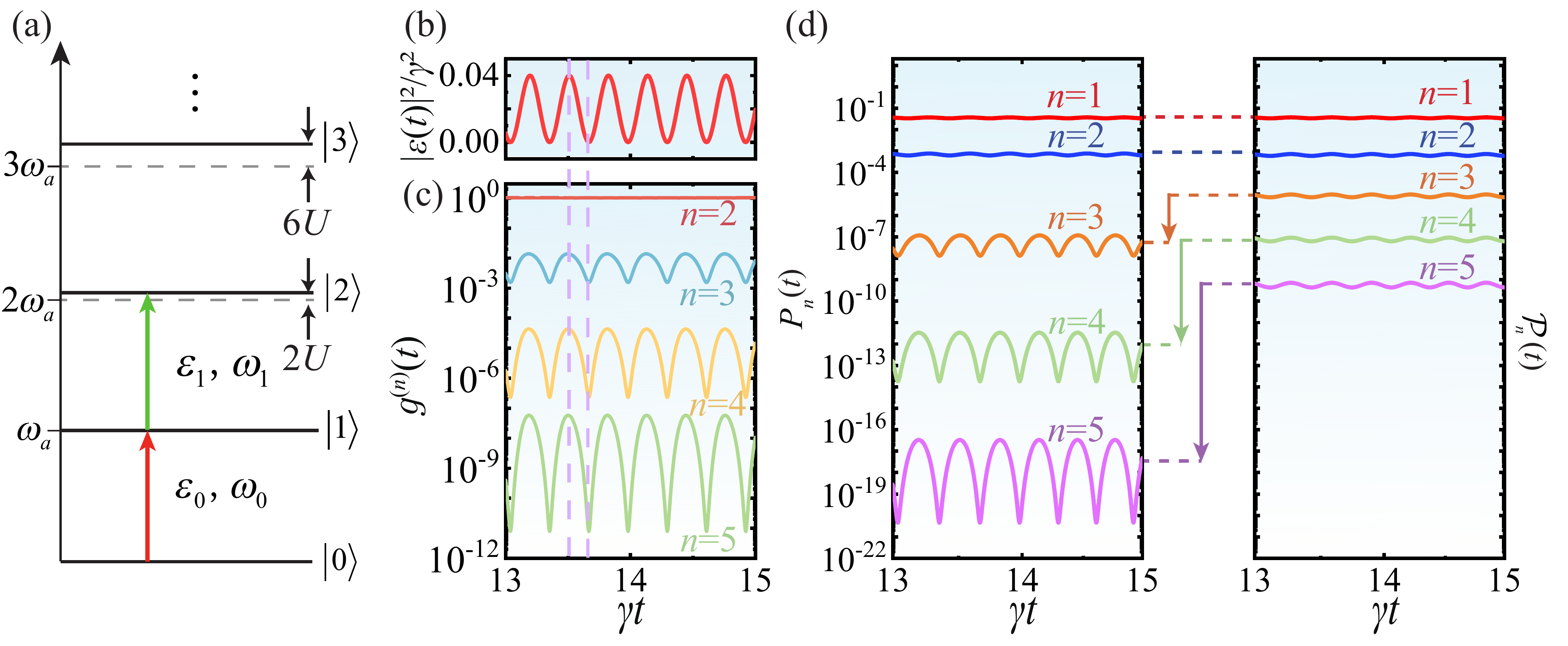}
		\caption{(Color online) (a) Energy level of the optical Kerr resonator. The optical resonator is driven by a two-tone coherent field with amplitudes ($\varepsilon _0$, $\varepsilon _1$) and frequencies ($\omega _0$, $\omega _1$), where the field ($\varepsilon _0$, $\omega_0$) is resonant to the transition $|0\rangle\rightarrow |1\rangle$ and the field ($\varepsilon _1$, $\omega_1$) is resonant to the transition $|1\rangle\rightarrow |2\rangle$. (b) The modulus square of the envelope of the driving field $|\varepsilon(t)/\gamma|^2= 0.04\sin^2(Ut)$ versus the scaled time $\gamma t$. (c) The equal-time $n$th-order correlation functions  $g^{(n)}(t)$ ($n=2,3,4,5$) versus $\gamma t$. (d) The photon-number distributions $P_{n}(t)$ and the Poisson distributions $\mathcal{P}_{n}(t)$ with the same mean photon number versus $\gamma t$.
			The parameters used are $U/\gamma=10$, $\varepsilon _0/\gamma=0.1$, $\varepsilon _1/\gamma=0.1$, $\Delta/\gamma=0$, and $\delta_1=2U$.}   
		\label{fig1}
	\end{figure*}
	
	\section{Physical model}\label{BIN}

We consider an optical Kerr resonator driven by a multi-tone field. The total system can be described by the Hamiltonian
\begin{equation}\label{eq1}
	H_{\rm sys}(t)=\hbar \omega_a a^{\dagger}a+\hbar Ua^{\dagger}a^{\dagger}aa+\hbar[\varepsilon (t)a+ \varepsilon^*(t)a^{\dagger} ],
\end{equation}
where $a^{\dagger}(a)$ is the creation (annihilation) operator of the optical mode with the resonance frequency $\omega_a$, $U = \hbar \omega_a^2cn_2/(n_1^2V_{\rm eff})$ is the Kerr nonlinear strength with the refractive index $n_1$, the nlinear refraction index $n_2$, the speed of light in vacuum $c$, and the effective mode volume $V_{\rm eff}$. 	
The Rabi frequency of the multi-tone drive is given by
	\begin{equation}\label{eq4}
		\varepsilon (t)= \sum_{n=0}^{N}\varepsilon_n e^{i\omega_nt},
	\end{equation}
where $\varepsilon_n=\sqrt{\gamma P_{{\rm in},n}/(\hbar \omega_n)}$ ($n=0,1,\cdots,N$) is the amplitude of the driving field with the frequency $\omega_n$ and power $P_{{\rm in},n}$, and $\gamma$ is the decay rate of the optical resonator.	

In the rotating frame with respect to $H_0=\omega_0 a^{\dagger}a $, the Hamiltonian becomes	
	\begin{equation}\label{eq4}
		H(t)=\hbar \Delta a^{\dagger}a+ \hbar Ua^{\dagger}a^{\dagger}aa+ \hbar [\varepsilon^{\prime}(t)a+\varepsilon^{\prime *}(t)a^{\dagger}],
	\end{equation}
where $\Delta\equiv \omega_a- \omega_0 $, and
\begin{equation}\label{eq5}
	\varepsilon^{\prime} (t)= \varepsilon_0+\sum_{n=1}^{N}\varepsilon_ne^{i\delta_nt},
\end{equation}
with $\delta_n\equiv \omega_n -\omega_0$  for $ n=1,2,...,N$.
As the total driving amplitude $\varepsilon^{\prime} (t)$ is time dependent, we will consider the dynamic behaviors of the system,
which is governed by the master equation~\cite{Carmichael1993}
\begin{equation}\label{MasterEq}
	\dot{\rho }=-\frac{i}{\hbar} \left[ H(t),\rho%
	\right]  +\frac{\gamma }{2}(2a\rho 
	a^{\dag }-a^{\dag }a\rho  -\rho  a^{\dag }a ).
\end{equation}
Here, $\rho$ is the density matrix of the system, and the master equation can be solved numerically by the open-source software QuTiP~\cite{QuTiP1,QuTiP2}.

There are two criteria mainly used to characterize the physical signature of the $n$PB effect~\cite{PhysRevA.87.023809}. The first criterion is based on the equal-time $n$th-order correlation function
\begin{equation}
	g^{(n)}\left( t\right) =
	\frac{\left\langle a^{\dag n}\left( t\right)a^n\left( t\right) \right\rangle}
	{\left\langle a^{\dag}\left( t\right)a\left( t\right) \right\rangle ^n}.
\end{equation}
For the single-photon transition process~\cite{PhysRevA.102.053710}, the correlation functions for the $n$PB effect ($n\geq2$) should be
\begin{equation}\label{eqgg}
g^{(n)}(t)\geqslant 1,\quad g^{(m)}(t)<1
\end{equation}
for $m>n$.
The other criterion is based on a comparison between the photon-number distributions $P_{n}(t)$ and the
Poisson distributions $\mathcal{P}_{n}(t)$ with the same mean photon number~\cite{PhysRevLett.118.133604}. The criterion for the $n$PB effect is defined by
\begin{equation}\label{eqpp}
P_{n}(t) \geqslant \mathcal{P}_{n}(t), \quad P_{m}(t)<\mathcal{P}_{m}(t),
\end{equation}
where $P_n(t)\equiv \left\langle n\right\vert \rho \left\vert n\right\rangle $ is the populations of Fock states for $n$ photons ($|n\rangle$, $n$ is a non-negative integer),
and $\mathcal{P}_{n}(t)=\langle\hat{n}(t)\rangle^{n} \exp (-\langle\hat{n}(t)\rangle) / n!$ is the Poisson distributions with the mean photon number $\left\langle \hat{n}(t) \right\rangle \equiv \left\langle a^{\dag }\left( t\right)a(t) \right\rangle$.
The criterion for the $n$PB effect in Eq.~(\ref{eqpp}) means that, in comparison with the coherent state, the probability of observing $n$ photons remains almost the same or increases, while the probabilities of detecting more than $n$ photons are suppressed.

The energy eigenstates of the optical Kerr resonator without external drive are the Fock states $|n\rangle$ ($n = 0, 1, 2, \dots$), with the corresponding eigen-energy
	\begin{equation}\label{eq22}
		E_n=\hbar \left(n+\frac{1}{2}\right)\omega_a+ \hbar Un(n-1).
	\end{equation}
The energy difference between the nearest-neighbor energy levels is given by
 \begin{equation}\label{eqEnd}
 	E_n-E_{n-1}=\hbar [ \omega_a+2(n-1)U],
 \end{equation}
which is a photon-number dependent variable.
Under the single-photon resonant driving $|0\rangle\rightarrow |1\rangle$, the anharmonic (photon-number dependent) energy levels will suppress the excitation of two-photon and other multiphoton states due to large detuning ($U \gg \gamma$). 
 
In this paper, we replace the single-tone drive by a multi-tone drive, and show that the $n$PB effect can be achieved by a $n$-tone drive with appropriate frequencies.
Specifically, when $\omega_0=\omega_a$, i.e., $\Delta=0$, and
\begin{equation}\label{eqdd2}
	\delta_n = 2(n-1)U,
\end{equation}
then the driving frequency $\omega_i$ ($i\leq n$) can match the transition $|i-1\rangle\rightarrow |i\rangle$ and all the transitions $|i-1\rangle\rightarrow |i\rangle$ are driven resonantly. But the excitation of the $(n+1)$th photon is still blockaded due to the anharmonicity of the energy spectrum. Thus, $n$PB effect can be observed with a $n$-tone drive.

In the following numerical calculations, we will select the experimentally accessible parameters ~\cite{vahala_optical_2003,spillane_ultrahigh-_2005,schuster_nonlinear_2008,noauthor_whispering_2011,huet_millisecond_2016,shen_compensation_2016,zielinska_self-tuning_2017,pavlov_soliton_2017}:
resonance wavelength $\lambda=1550$ nm, the quality factor $Q=\omega_a/\gamma=2.5 \times 10^9$, $V_{\rm eff}=196$ $\mu$m$^3$, $n_1=1.4$, and $n_2=4 \times 10^{-14}$ m$^2$/W. In the microring resonators, the quality factor $Q$ can reach $10^9-10^{12}$~\cite{noauthor_whispering_2011,huet_millisecond_2016,pavlov_soliton_2017}, and $V_{\rm eff}$ is typically $10^2-10^4$ $\mu$m$^3$~\cite{vahala_optical_2003,spillane_ultrahigh-_2005}. The Kerr coefficient can be $n_2\sim 10^{-14}$ for materials with potassium titanyl phosphatep~\cite{zielinska_self-tuning_2017}.

\section{Two-photon Blockade}\label{CIN}

In this section, we consider the two-photon blockade case in which the optical resonator is driven by a two-tone coherent field. The envelope of the driving field reads
$|\varepsilon(t)| = |\varepsilon_0+\varepsilon_1e^{i\delta_1t}|$,
where $\varepsilon _0$ and $\varepsilon _1$ denote the amplitudes of the driving fields with frequencies $\omega_0$ and $\omega_1$, respectively.
As mentioned above, we will consider the two-photon resonance conditions $\Delta=0$ and $\delta_1=2U=20\gamma$, as shown in Fig.~\ref{fig1}(a).
Under the weak driving condition ($\varepsilon_0=\varepsilon_1=0.1 \gamma$), the modulus square of the envelope of the driving field is given by $|\varepsilon(t)/\gamma|^2= 0.04\sin^2(Ut)$ [see Fig.~\ref{fig1}(b)].
It oscillates with the maximal amplitude $0.04 \gamma$ and period $T =\pi/U$.

\begin{figure}[htbp]
	\centering
	\includegraphics[bb=70 59 940 785, width=8.5 cm, clip]{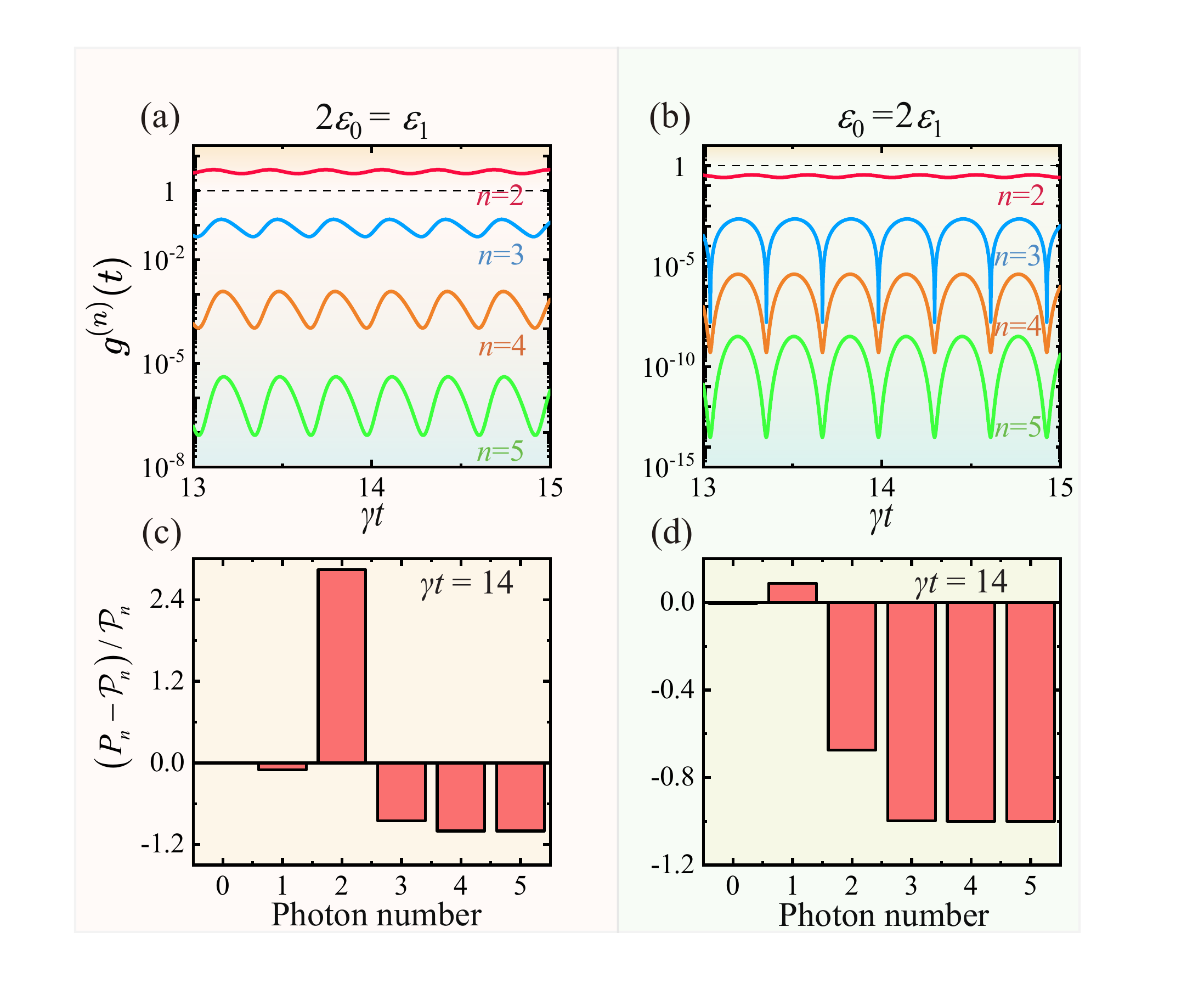}
	\caption{(Color online) (a) and (b) The time evolution of the equal-time $n$th-order correlation functions  $g^{(n)}(t)$ ($n=2,3,4,5$) for different driving amplitudes: (a) $\varepsilon _0/\gamma=0.1, \varepsilon _1/\gamma=0.2$; (b) $\varepsilon _0/\gamma=0.2, \varepsilon _1/\gamma=0.1$. (c) and (d) The deviations of the photon distribution $P_{n}(t)$ from the standard Poisson distribution $\mathcal{P}_{n}(t)$ with the same mean photon number at $\gamma t=14$ for different driving amplitudes: (c) $\varepsilon _0/\gamma=0.1, \varepsilon _1/\gamma=0.2$; (d) $\varepsilon _0/\gamma=0.2, \varepsilon _1/\gamma=0.1$.
		Other parameters used are $U/\gamma=10$, $\Delta/\gamma=0$, and $\delta_1=2U$.}   
	\label{fig2}
\end{figure}

\begin{figure}[htbp]
	\centering
	\includegraphics[bb=46 62 1722 702, width=8.5 cm, clip]{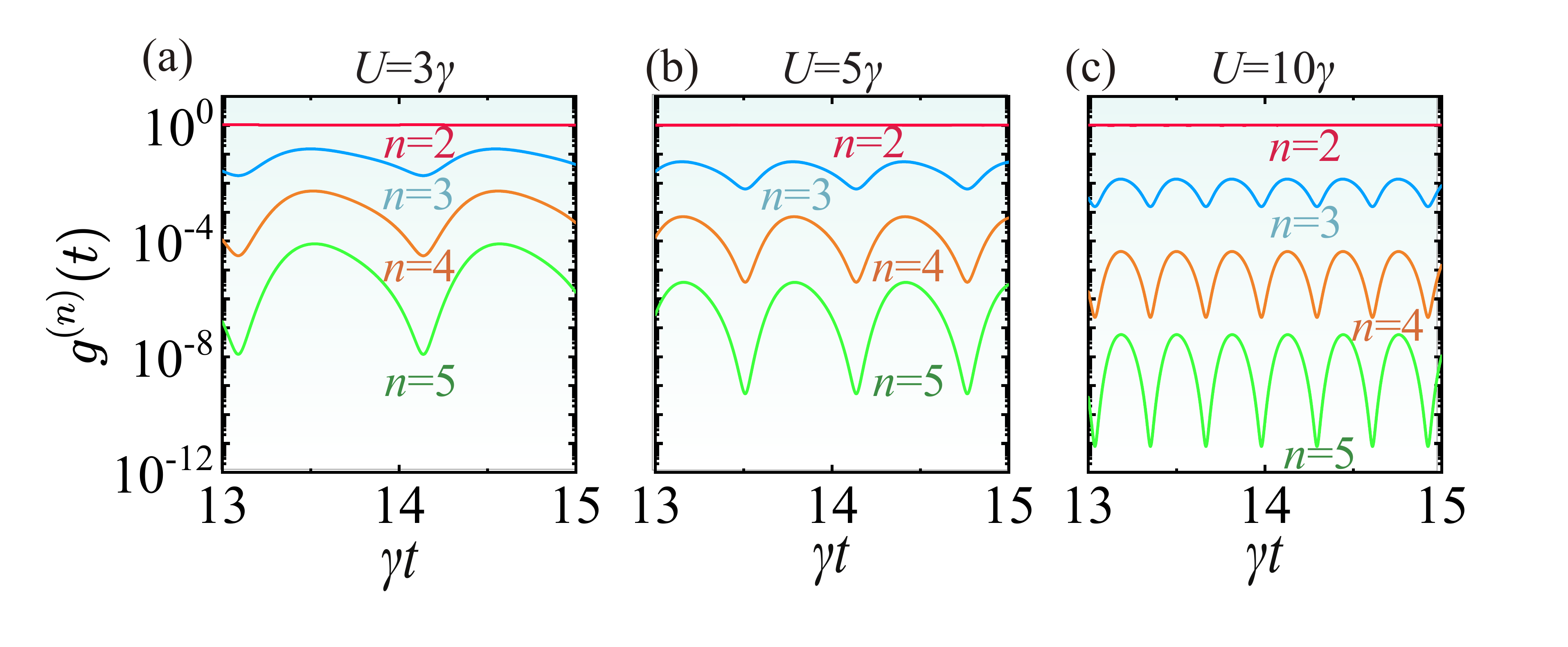}
	\caption{(Color online) The time evolution of the equal-time $n$th-order correlation functions $g^{(n)}(t)$ ($n=2,3,4,5$) are plotted for different values of the scaled Kerr nonlinear strength $U/\gamma$: (a) $U/\gamma=3$; (b) $U/\gamma=5$; (c) $U/\gamma=10$.
		Other parameters used are $\varepsilon _0/\gamma=0.1$, $\varepsilon _1/\gamma=0.1$, $\Delta/\gamma=0$, and $\delta_1=2U$. }   
	\label{fig3}
\end{figure}

The equal-time $n$th-order correlation functions $g^{(n)}(t)$ ($n=2,3,4,5$) are shown in Fig.~\ref{fig1}(c). 
We find that $g^{(2)}(t)\approx 1$, and $g^{(n)}(t)\ll 1$ for $(n=3,\;4,\;5)$, which indicate that the 2PB is achieved and the probabilities of Fock states with more than two photons are suppressed.
As the amplitude of the driving field $\varepsilon(t)$ is time-dependent, the correlation functions $g^{(n)}(t)$ also oscillate with the same period $T =\pi/U$ as $|\varepsilon(t)/\gamma|^2$.
Importantly, the criteria for 2PB is always satisfied in the stationary regime, i.e., $g^{(2)}(t)\geqslant 1$ and $g^{(m)}(t)<1$ ($m\geq 3$) for $t\gg1/\gamma$, which is different from the previous results about the dynamic PB~\cite{PhysRevA.93.043857,PhysRevA.90.013839,PhysRevA.90.063805,PhysRevB.97.241301,PhysRevLett.123.013602,PhysRevLett.129.043601,LIU2024,Geng:24,PhysRevA.110.023718} that appears periodically in pulses.

Furthermore, we compare the photon-number distributions $P_{n}(t)$ and the Poisson distributions $\mathcal{P}_{n}(t)$ with the same mean photon number in Fig.~\ref{fig1}(d).
We can see that the criterion in Eq.~(\ref{eqpp}) for the $n$PB effect is satisfied in the stationary regime ($t\gg1/\gamma$): $P_{1}(t) \approx  \mathcal{P}_{1}(t)$ and $P_{2}(t) \approx  \mathcal{P}_{2}(t)$, while $ P_{3}(t)\ll\mathcal{P}_{3}(t)$, $P_{4}(t)\ll\mathcal{P}_{4}(t)$, and $P_{5}(t)\ll\mathcal{P}_{5}(t)$. 
This phenomenon can be understood as follows. The transitions $|0\rangle\rightarrow |1\rangle$ and  $|1\rangle\rightarrow |2\rangle$ are driven resonantly, as shown in Fig.~\ref{fig1}(a), while the transitions $|2\rangle\rightarrow |3\rangle$, $|3\rangle\rightarrow |4\rangle$, and $|4\rangle\rightarrow |5\rangle$ are largely detuned from the driving field.

Interestingly, the correlation functions $g^{(n)}(t)$ and the photon-number distributions $P_{n}(t)$ can be manipulated by tuning the relative amplitudes $\varepsilon_1$ and $\varepsilon_2$.
As shown in Figs.~\ref{fig1}(c) and \ref{fig1}(d), we have $g^{(2)}(t)\approx 1$ and $P_{1}(t)/\mathcal{P}_{1}(t)\approx P_{2}(t)/\mathcal{P}_{2}(t)\approx1$ for $\varepsilon_1=\varepsilon_2$. 
Differently, if $\varepsilon _1$ is greater than $\varepsilon _0$, as shown in Figs.~\ref{fig2}(a) and \ref{fig2}(c), then we have $g^{(2)}(t) > 1$, $P_{1}(t)/\mathcal{P}_{1}(t)<1$, and $P_{2}(t)/\mathcal{P}_{2}(t)>1$, which indicate that the transition $|1\rangle\rightarrow |2\rangle$ is enhanced for $\varepsilon _1>\varepsilon _0$. 
In contrast, if $\varepsilon _1<\varepsilon _0$, then all the correlation functions $g^{(n)}(t)$ are less than 1, as shown in Fig.~\ref{fig2}(b).
These mean that, in comparison with the Poisson distributions $\mathcal{P}_{n}(t)$, the two-photon probability $P_{2}(t)$ is still suppressed for $\varepsilon _1<\varepsilon _0$, as shown in Fig.~\ref{fig2}(d).
Therefore, in order to achieve 2PB by a two-tone driving field, we should set $\varepsilon _1\geq \varepsilon _0$.

	\begin{figure}[htbp]
		\includegraphics[bb=2 6 300 300, width=8.5 cm, clip]{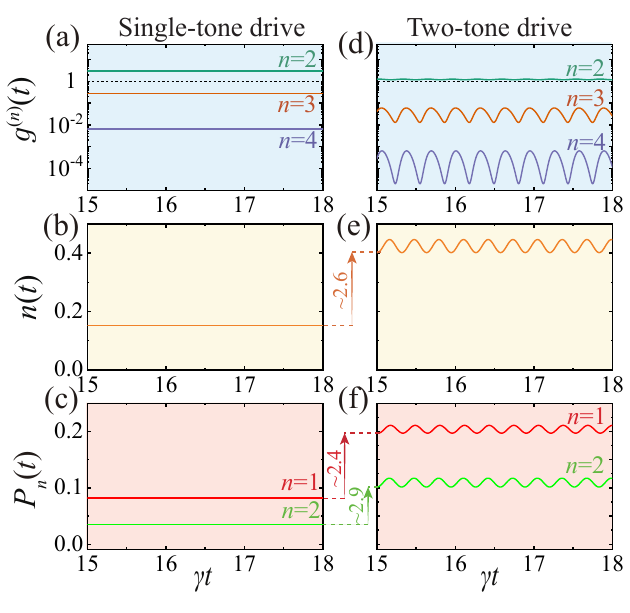}
		\caption{(Color online) (a) and (d) The equal-time $n$th-order correlation functions  $g^{(n)}(t)$ ($n=2,3,4$) versus the scaled time $\gamma t$. (b) and (e) The mean photon number $n(t)$ versus $\gamma t$. (c) and (f) The photon-number distributions $P_{n}(t)$ versus $\gamma t$. (a)-(c) The resonator is driven by a single-tone drive, i.e., $\omega_0=\omega_1=\omega_a+U$ and $\varepsilon_0+\varepsilon_1=1.2\gamma $; (d)-(f) the resonator is driven by a two-tone drive with the parameters $\omega_0=\omega_a$, $\omega_1=\omega_a+2U$, $\varepsilon_0=0.5\gamma $, and $\varepsilon_1=0.7\gamma $. The Kerr nonlinear strength is $U=10\gamma$.} 
		\label{fig4}
	\end{figure}

	\begin{figure*}[htbp]
		\includegraphics[bb=8 29 1885 800, width=14 cm, clip]{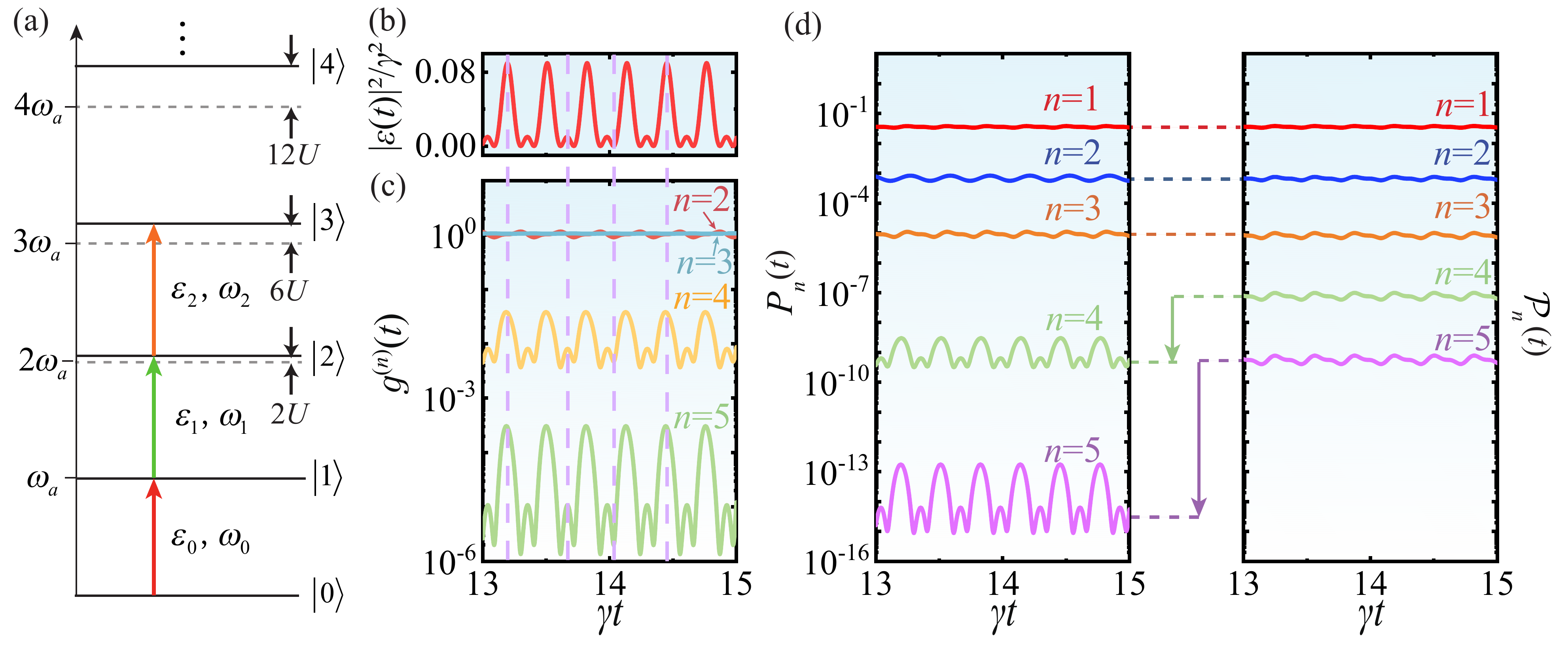}
		\caption{(Color online) (a) Energy level of the optical Kerr resonator and the three-tone coherent field with amplitudes ($\varepsilon _0$, $\varepsilon _1$, $\varepsilon _2$) and frequencies ($\omega _0$, $\omega _1$, $\omega _2$), where the field ($\varepsilon _0$, $\omega_0$) is resonant to the transition $|0\rangle\rightarrow |1\rangle$, the field ($\varepsilon _1$, $\omega_1$) is resonant to the transition $|1\rangle\rightarrow |2\rangle$, and the field ($\varepsilon _2$, $\omega_2$) is resonant to the transition $|2\rangle\rightarrow |3\rangle$. (b) The modulus square of the amplitude $|\varepsilon(t)/\gamma|^2$ of the driving field versus the scaled time $\gamma t$. (c) The equal-time $n$th-order correlation functions  $g^{(n)}(t)$ ($n=2,3,4,5$) versus $\gamma t$. (d) The photon-number distributions $P_{n}(t)$ and the Poisson distributions $\mathcal{P}_{n}(t)$ of a coherent state with the same mean photon number versus $\gamma t$.
			The parameters used are $U/\gamma=10$, $\varepsilon _0/\gamma=0.1$, $\varepsilon _1/\gamma=0.1$, $\varepsilon _2/\gamma=0.1$, $\Delta/\gamma=0$, $\delta_1=2U$, and $\delta_2=4U$.} 
		\label{fig5}
	\end{figure*}

We know that the Kerr nonlinear strength $U$ is a significant parameter for the appearance of PB.
The correlation functions $g^{(n)}(t)$ are plotted for different values of the scaled Kerr nonlinear strength $U/\gamma$ in Fig.~\ref{fig3}.
Here, we can find two features: (i) The blockade effect becomes notable with the increase of the nonlinear strength $U$.
(ii) The oscillation frequency of the correlation functions $g^{(n)}(t)$ becomes higher with the increase of $U$. That is because the modulus square of the amplitude $|\varepsilon(t)/\gamma|^2$ of the driving field oscillates with the period $T =\pi/U$.
It is worth mentioning that the 2PB can be observed with a modest Kerr nonlinear strength $U=3\gamma$.

To show one of the advantages of the multi-tone drive over the single-tone drive, we compare the results corresponding to the two drives in Fig.~\ref{fig4}.
As mentioned in some previous works~\cite{PhysRevA.87.023809,PhysRevA.90.013839}, two-photon blockade under a single-tone drive can only be observed with a strong driven field. In Fig.~\ref{fig4}(a), we take a driving strength $\varepsilon_0+\varepsilon_1=1.2\gamma $ for single-tone drive with $\omega_0=\omega_1=\omega_a+U$, to obtain two-photon blockade, i.e., $g^{(2)}(t)>1$, $g^{(3)}(t)<1$, and $g^{(4)}(t)<1$.
Correspondingly, two-photon blockade also appears under a two-tone drive with the frequencies $\omega_0=\omega_a$ and $\omega_1=\omega_a+2U$, and strengthes $\varepsilon_0=0.5\gamma $ and $\varepsilon_1=0.7\gamma $, as shown in Fig.~\ref{fig4}(d).
Most interestingly, under the same driving strength $\varepsilon_0+\varepsilon_1=1.2\gamma$ with the frequency $\omega_0+\omega_1=2\omega_a+2U$, the mean photon $n(t)$ under the two-tone drive is about $2.6$ times that of the case under single-tone drive, as shown in Figs.~\ref{fig4}(b) and \ref{fig4}(e).
To show more details, the photon distributions $P_1(t)$ and $P_2(t)$ for both the single-tone and two-tone drives are shown in Figs.~\ref{fig4}(c) and \ref{fig4}(f), respectively. We can see that $P_1(t)$ [$P_2(t)$] by the two-tone drive is about 2.4 (2.9) times that of the case by the single-tone drive, and $P_2(t)$ by the two-tone drive is even larger than $P_1(t)$ by the single-tone drive.
The enhancement of the photon generation by multi-tone drive is even more evident under the weak driving condition.

	\section{Three-photon Blockade}\label{DIN}

In this section, we demonstrate that the 3PB can be achieved by driving the Kerr optical resonator with a three-tone field. The driving amplitude for a three-tone field is given by $\varepsilon (t) = \varepsilon_0+\varepsilon_1e^{i\delta_1t}+\varepsilon_2e^{i\delta_2t}$.	
The detunings are $\delta_1=2U$ and $\delta_2=4U$ for the three-photon resonant driving, and the energy-level diagram of the system is shown in Fig.~\ref{fig5}(a). 
Under the weak-driving conditions $\varepsilon_0/\gamma=\varepsilon_1/\gamma=\varepsilon_2/\gamma=0.1$, the modulus square of the scaled driving amplitude $|\varepsilon(t)/\gamma|^2$ versus the scaled time $\gamma t$ is shown in Fig.~\ref{fig5}(b).
It also oscillates with a period $T =\pi/U$, and there are two peaks in one period with the maximal amplitude $0.09\gamma$.
Corresponding to the driving amplitude $\varepsilon(t)$, the correlation functions  $g^{(n)}(t)$ ($n=2,3,4,5$) are plotted as functions of the scaled time $\gamma t$ in Fig.~\ref{fig5}(c). 
We can see that, in the stationary regime of $t\gg1/\gamma$, both $g^{(2)}(t)$ and $g^{(3)}(t)$ oscillate around $1$ with a very small amplitude, while $g^{(4)}(t)$ and $g^{(5)}(t)$ are much smaller than $1$, indicating that the 3PB appears in the system.

	\begin{figure}[htbp]
	\includegraphics[bb=145 3 1685 985, width=8.5 cm, clip]{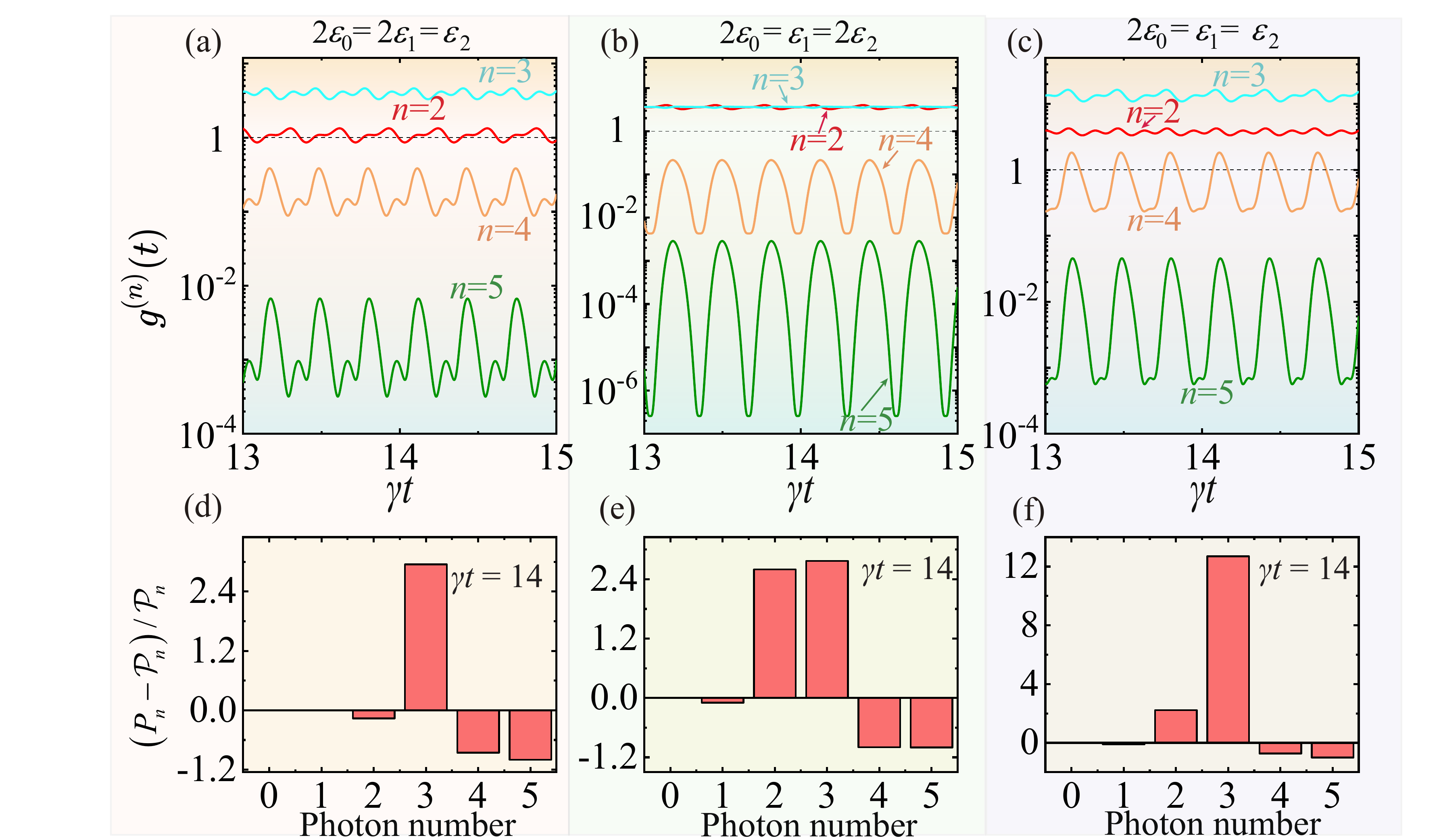}
	\caption{(Color online) (a)-(c) The time evolution of the equal-time $n$th-order correlation functions  $g^{(n)}(t)$ ($n=2,3,4,5$) for different driving amplitudes: (a) $\varepsilon _0/\gamma=0.1$, $\varepsilon _1/\gamma=0.1$, $\varepsilon _2/\gamma=0.2$; (b) $\varepsilon _0/\gamma=0.1$, $\varepsilon _1/\gamma=0.2$, $\varepsilon _2/\gamma=0.1$; (c) $\varepsilon _0/\gamma=0.1$, $\varepsilon _1/\gamma=0.2$, $\varepsilon _2/\gamma=0.2$. (d)-(f) The deviations of the photon distribution $P_{n}(t)$ from the standard Poisson distribution $\mathcal{P}_{n}(t)$ with the same mean photon number at $\gamma t=14$ for different driving amplitudes: (d) $\varepsilon _0/\gamma=0.1$, $\varepsilon _1/\gamma=0.1$, $\varepsilon _2/\gamma=0.2$; (e) $\varepsilon _0/\gamma=0.1$, $\varepsilon _1/\gamma=0.2$, $\varepsilon _2/\gamma=0.1$; (f) $\varepsilon _0/\gamma=0.1$, $\varepsilon _1/\gamma=0.2$, $\varepsilon _2/\gamma=0.2$. Other parameters used are $U/\gamma=10$, $\Delta/\gamma=0$, $\delta_1=2U$ and $\delta_2=4U$.} 
	\label{fig6}
\end{figure}

In addition, the photon-number distributions $P_{n}(t)$ and the Poisson distributions $\mathcal{P}_{n}(t)$ with the same mean photon number versus the scaled time $\gamma t$ are shown in Fig.~\ref{fig5}(d). We find $P_{1} \approx  \mathcal{P}_{1}$, $P_{2} \approx  \mathcal{P}_{2}$, $P_{3}\approx {P}_{3}$, $P_{4}<\mathcal{P}_{4}$, and $P_{5}<\mathcal{P}_{5}$ in the stationary regime ($t\gg1/\gamma$). According to the criterion in Eq.~(\ref{eqpp}), a 3PB can be observed in the system under the three-tone drive.

	\begin{figure}[htbp]
	\includegraphics[bb=2 6 1838 756, width=8.5 cm, clip]{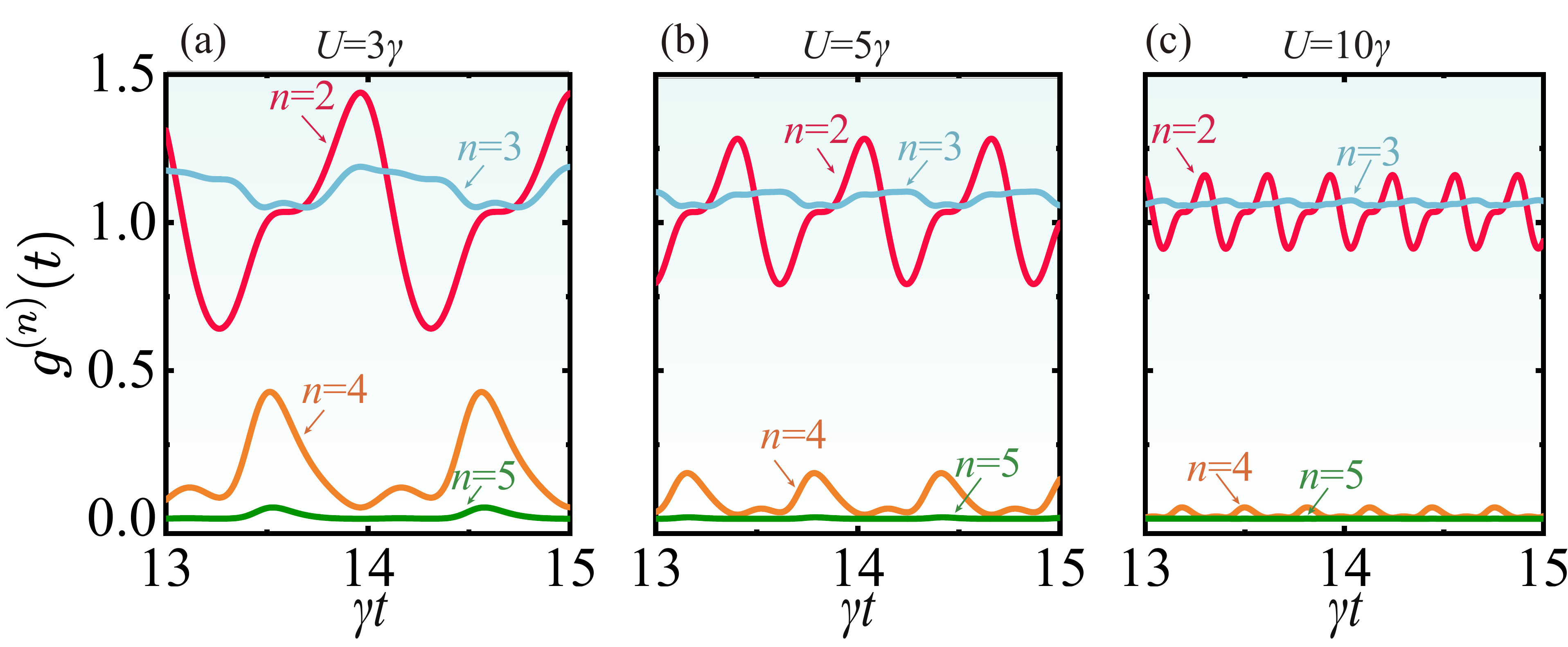}
	\caption{(Color online) The time evolution of the equal-time $n$th-order correlation functions $g^{(n)}(t)$ ($n=2,3,4,5$) are plotted for different values of the scaled Kerr nonlinear strength $U/\gamma$: (a) $U/\gamma=3$; (b) $U/\gamma=5$; (c) $U/\gamma=10$.
		Other parameters used are $\varepsilon _0/\gamma=0.1$, $\varepsilon _1/\gamma=0.1$, $\varepsilon _2/\gamma=0.1$, $\Delta/\gamma=0$, $\delta_1=2U$, and $\delta_2=4U$.} 
	\label{fig7}
\end{figure}

We also discuss the effect of the driving amplitudes ($\varepsilon _0$, $\varepsilon _1$, and $\varepsilon _2$) on the 3PB.
As discussed in the previous section, we need $\varepsilon _1\geq \varepsilon _0$ to achieve the 2PB.
Now, there are three driving amplitudes ($\varepsilon _0$, $\varepsilon _1$, and $\varepsilon _2$), and we set $\varepsilon _1\geq \varepsilon _0$ to make sure that the excitation of the second photon is not blockaded.
So there are three cases, as shown in Fig.~\ref{fig6}: (a) $\varepsilon _0=\varepsilon _1<\varepsilon _2$, we have $g^{(2)}(t)\approx 1$, $g^{(3)}(t)>1$, $g^{(4)}(t)<1$, $g^{(5)}(t)<1$; (b) $\varepsilon _1>\varepsilon _0=\varepsilon _2$, we have $g^{(2)}(t)\approx g^{(3)}(t)>1$, $g^{(4)}(t)<1$, $g^{(5)}(t)<1$; (c) $\varepsilon _0<\varepsilon _1=\varepsilon _2$, we have $g^{(3)}(t)>g^{(2)}(t)>1$, $g^{(5)}(t)<1$, $g^{(4)}(t)$ oscillates around $1$.
The deviations of the photon distribution $P_{n}(t)$ from the standard Poisson distribution $\mathcal{P}_{n}(t)$ indicate that 
the 3PB appears in all the three cases.

Under the driving of a three-tone driving field, the correlation functions $g^{(n)}(t)$ for the photons in the optical resonator are plotted for different values of the scaled Kerr nonlinear strength $U/\gamma$ in Fig.~\ref{fig7}. It can be observed that the system already exhibits 3PB in the case of $U=3\gamma$. Moreover, as the nonlinear strength increases, the oscillation amplitudes of these correlation functions decrease gradually, indicating a more stable and pronounced photon blockade effect.

\section{Conclusions}\label{Con}
In conclusion, we have proposed a scheme to achieve multiphoton blockade effect under the multi-tone drive. Concretely, we have demonstrated that both the 2PB and 3PB can be achieved by driving an optical Kerr resonator with the two-tone and three-tone driving field, respectively.
Under the multi-tone drives, though there is no steady-state solution of the model, we can continuously achieve multiphoton blockade in the stationary regime, even with a modest Kerr nonlinear strength. 
The multi-tone drive scheme also provides an efficient way to manipulate the correlation functions and photon distributions by tuning the relative amplitudes of the lasers with different frequencies.
In principle, we can extend the scheme to achieve $n$-photon blockades by a $n$-tone drive with specific resonance frequencies. 
The scheme is experimentally accessible with the state-of-art conditions, and it can be extended to other bosonic systems, such as phononic~\cite{PhysRevA.93.013808,PhysRevA.82.032101,xu_phonon_2016,shi_phonon_2019} and magnonic~\cite{PhysRevB.100.134421,wang_magnon_2020,PhysRevA.101.042331,PhysRevA.106.013705,PhysRevA.101.063838} systems.

\begin{acknowledgments}
X.W.X. is supported by the Innovation Program for Quantum Science and Technology (Grant No.~2024ZD0301000), the Science and Technology Innovation Program of Hunan Province (Grant No.~2022RC1203), the National Natural Science Foundation of China (NSFC) (Grants No.~12064010, No.~12247105, and No.~12421005), and Hunan provincial major sci-tech program (Grant No.~2023ZJ1010). J.-Q.L. is supported in part by NSFC (Grants No.~12175061, No.~12247105, No.~11935006, and No.~12421005), National Key Research and Development Program of China (Grant No.~2024YFE0102400), and Hunan Provincial Major SciTech Program (Grant No.~2023ZJ1010).
\end{acknowledgments}

	\bibliography{ref2}

\end{document}